\documentclass[journal,twoside]{IEEEtran}

\usepackage{xcolor,soul,framed} 

\colorlet{shadecolor}{yellow}
\usepackage[pdftex]{graphicx}
\graphicspath{{../pdf/}{../jpeg/}}
\DeclareGraphicsExtensions{.pdf,.jpeg,.png}

\usepackage[cmex10]{amsmath}
\usepackage{mathabx}
\usepackage{array}
\usepackage{mdwmath}
\usepackage{mdwtab}
\usepackage{eqparbox}
\usepackage{url}
\usepackage{cite}
\usepackage{multirow}
\usepackage{makecell}
\usepackage{subcaption}
\usepackage{algorithm}
\usepackage{algpseudocode}
\usepackage{pgffor}
\usepackage{eso-pic}
\usepackage{tabularx}  
\usepackage{booktabs} 

\newcommand{\AcceptedNotice}{%
  \AddToShipoutPictureBG*{%
    \AtPageUpperLeft{%
      \put(\LenToUnit{0.5\paperwidth},\LenToUnit{-1.0cm}){%
        \makebox[0pt][c]{\parbox{0.85\paperwidth}{\centering\footnotesize
          Accepted for publication at the 2027 IEEE PES Grid Edge Technologies
          Conference \& Exposition, Salt Lake City, UT, USA, April 19--22, 2027.}}}}%
    \AtPageLowerLeft{%
      \put(\LenToUnit{0.5\paperwidth},\LenToUnit{0.8cm}){%
        \makebox[0pt][c]{\parbox{0.85\paperwidth}{\centering\scriptsize
          \copyright~2027 IEEE. Personal use of this material is permitted.
          Permission from IEEE must be obtained for all other uses, in any current
          or future media, including reprinting/republishing this material for
          advertising or promotional purposes, creating new collective works, for
          resale or redistribution to servers or lists, or reuse of any copyrighted
          component of this work in other works.}}}}%
  }%
}

\begin{document}
\title{Impact of Grid-Interactive Data Center Microgrids on Grid Reliability\\
}

 \author{Kiran Kumar Challa, \IEEEmembership{Senior Member, IEEE}, and Kumarsinh Jhala, \IEEEmembership{Senior Member, IEEE.}  
 \thanks{K. K. Challa and Kumarsinh Jhala are with Argonne National Laboratory, Lemont, IL, USA
(e-mail: kchalla@anl.gov, kjhala@anl.gov).}
 \thanks{The submitted manuscript has been created by UChicago Argonne, LLC, Operator of Argonne National Laboratory (“Argonne”). Argonne National Laboratory's work was supported by the U.S. Department of Energy, Office of Electricity, under contract DE-AC02-06CH11357. The U.S. Government retains for itself, and others acting on its behalf, a paid-up nonexclusive, irrevocable worldwide license in said article to reproduce, prepare derivative works, distribute copies to the public, and perform publicly and display publicly, by or on behalf of the Government.}%
}


\AcceptedNotice
\maketitle


\begin{abstract}
Traditional data centers, operating as passive loads with standby backup resources, present significant challenges for grid integration due to their scale, rapid load fluctuations, and limited operational flexibility. This article investigates the transformative potential of grid-interactive data center microgrids, which integrate battery energy storage, on-site generation, and intelligent microgrid controllers to actively support both facility and grid operations. The proposed simulation framework enables the study comparing conventional and grid-interactive architectures under a range of steady-state and contingency scenarios. Results reveal that grid-interactive microgrids can substantially reduce peak grid demand, improve voltage stability, and improve system reliability—especially when data center and system load peaks coincide. The findings highlight the critical role of coordinated resource management in enabling the reliable and efficient integration of large data center loads, and provide a practical methodology for assessing their system-level impacts.

\end{abstract}

\begin{IEEEkeywords}
Data center microgrids, large-load integration, battery energy storage systems, power system reliability.
\end{IEEEkeywords}

\section{Introduction}
\IEEEPARstart {T}The rapid expansion of cloud computing, artificial intelligence (AI), high-performance computing (HPC), and digital services is driving unprecedented growth in data center electricity demand \cite{ieee_grid_flexibility}. Recent studies project that data centers could account for 9--17\% of total U.S. electricity consumption by 2030. Furthermore, global data center electricity consumption is expected to approximately double by 2030, making data centers one of the fastest-growing categories of electrical demand in modern power systems \cite{iea_ai_energy}. The increasing concentration of large data center campuses, many exceeding hundreds of megawatts and approaching gigawatt-scale capacities, has created significant challenges for utilities, transmission planners, and system operators.

Unlike traditional industrial loads, modern data centers exhibit unique electrical characteristics. Large fractions of the facility demand are supplied through power-electronic interfaces including server power supplies, UPS systems, battery storage systems, and inverter-based energy resources \cite{sheng2026_ai_datacenter}. In addition, AI workloads can introduce rapid load variations, high power densities, and significant cooling requirements that directly affect system planning, operational flexibility, and reliability assessments \cite{dc_stability_risks}. Recent reliability assessments by the North American Electric Reliability Corporation (NERC) have identified large computational loads as an emerging reliability concern due to their scale, fast-changing demand characteristics, and potential interactions with grid disturbances \cite{nerc_datacenter_load_modeling}. 

Most existing data centers employ a reliability-centric electrical architecture in which the utility grid supplies the facility during normal operation, while UPS systems and standby diesel generators provide ride-through and backup power during utility outages \cite{UptimeTierStandard,SchneiderRD101,VertivPowerTrain2025}. Although this approach satisfies the high availability requirements of mission-critical facilities, the backup resources remain largely idle during normal operation, causing the bulk power system to treat the facility as a passive load with limited operational flexibility. As data center capacities continue to grow, there is increasing interest in utilizing these on-site energy resources to support grid services and improve utility interactions \cite{nerc_datacenter_load_modeling}.

High reliability concerns and increasing interconnection have motivated the development of grid-interactive data center architectures to enable the faster interconnection approvals. In these architectures, battery energy storage systems (BESS), distributed grid resources (DGRs), inverter-based resources (IBRs), and intelligent microgrid controllers are actively coordinated to manage facility demand and support grid operations \cite{grid_interactive_datacenters}. Rather than functioning solely as backup resources, these assets can participate in peak demand reduction, frequency response, voltage regulation, demand response, congestion management, and controlled islanding operations \cite{8340204}. Such capabilities enable data centers to transition from passive electrical loads to active grid participants capable of enhancing both facility resilience and system reliability.

Recent studies have developed simulation frameworks for evaluating conventional and grid-interactive data center architectures under representative operating scenarios. The operational cost of geo-distributed data center microgrids is optimized in \cite{Yu2018} through coordinated energy management. An optimization methodology for battery sizing and charging/discharging schedules is presented in \cite{Thompson2016} to maximize economic benefits for data center owners. A power-sharing framework for interconnected data center microgrids is proposed in \cite{Aghmadi2025} to improve resilience and reduce reliance on fossil-fuel backup generation. Furthermore, a controller-hardware-in-the-loop (C-HIL) digital twin of a data center is developed in \cite{Talukdar2026} to demonstrate the transient support that data centers can provide to the power system.

While these studies have significantly advanced data center energy management, resource sizing, and control strategies, they primarily focus on improving facility-level operational performance and economics. Comparatively less attention has been given to systematically evaluating how the inherent flexibility of grid-interactive data center microgrids influences bulk power system reliability under diverse operating conditions. In particular, the impacts of coordinated operation on voltage performance, contingency response, peak-load conditions, and varying temporal alignment between data center demand and system loading remain insufficiently quantified. Addressing these aspects is essential for understanding the evolving role of large data centers as active grid assets and for quantifying the system-level benefits that grid-interactive operation can provide to utilities and transmission planners. The Following contributions address these research gaps:

\begin{enumerate}
    \item Develops a comparative assessment methodology for evaluating conventional and grid-interactive data centers across different load-alignment and contingency conditions.

    \item Quantifies the grid impacts of grid-interactive data centers in terms of peak-load reduction, minimum-voltage improvement, and post-contingency voltage security.

    \item Quantifies how temporal alignment between data-center and system demand determines the effectiveness of local generation and BESS in mitigating system peak loading and voltage violations.

\end{enumerate}
These contributions provide a practical methodology for assessing the transition of large data centers from passive loads to grid-interactive microgrids for planning and operational studies.
\section{Data Center Power Architectures}
From a power-system perspective, data center architectures can be broadly categorized based on how their on-site resources interact with the utility grid. Two primary architectures are considered in this study. The first represents the traditional architecture in which the utility grid serves the facility during normal operation, while backup batteries and diesel generators remain in standby mode and are utilized primarily during utility outages. The second consists of grid-interactive data center microgrids that actively coordinate distributed energy resources, battery energy storage systems, and controllable loads to support both facility reliability and grid operations. Since these architectures exhibit substantially different operational characteristics, their representations in bulk power system studies must also differ.

\subsubsection{Conventional Data Center with Standby Backup Resources}

\begin{figure}[!h]
\centering
\includegraphics[width=\linewidth]{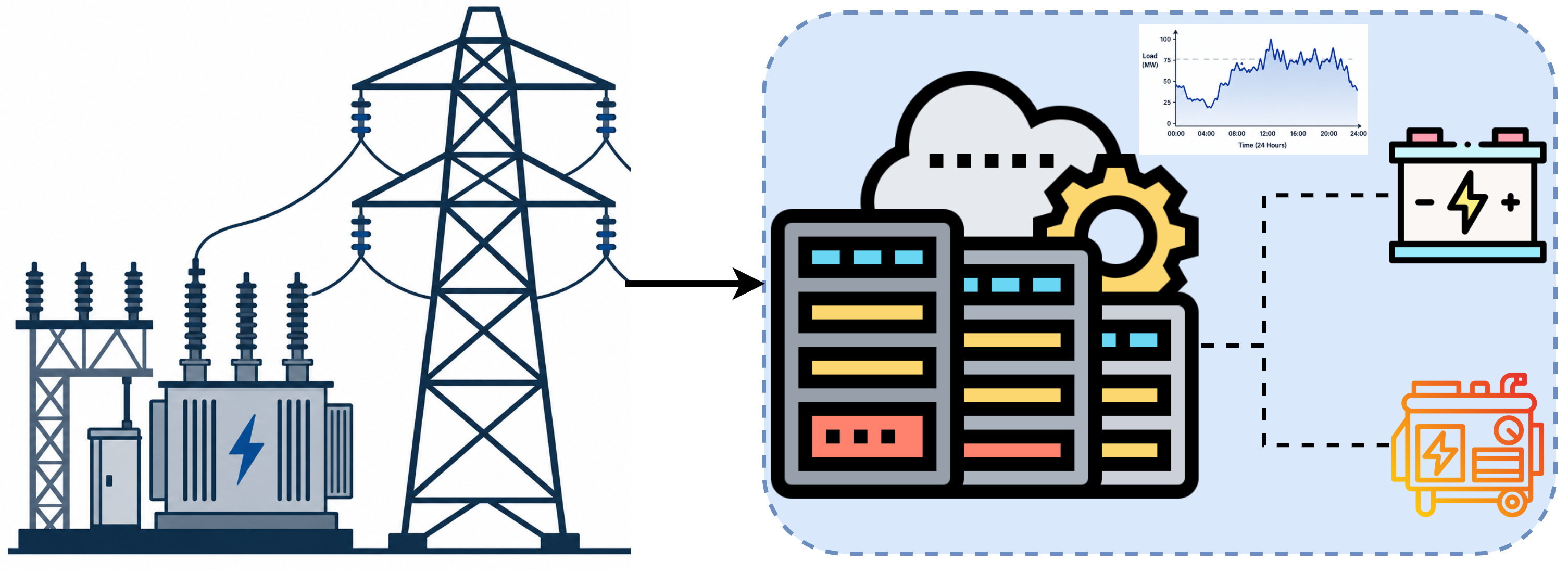}
\caption{Conventional data center architecture.}
\label{fig:dc_backup_architecture}
\end{figure}

The majority of existing data centers employ a conventional architecture where the utility grid serves the facility during normal operation, and backup resources remain inactive until a utility outage occurs. As shown in Fig.~\ref{fig:dc_backup_architecture}, the primary objective of this architecture is maintaining high availability through redundant power supply paths rather than providing operational flexibility to the electric grid.

The architecture consists of three primary components: primary utility supply, backup power generation, and critical loads. During normal operating conditions, the utility supplies nearly all facility demand while the backup generators contribute zero active power output. The UPS battery system provides short-duration ride-through capability, bridging the transition period required for diesel generator startup and synchronization. Consequently, the bulk power system primarily observes the facility as a large, nearly constant-power load with limited interaction beyond power consumption.

Following a utility disturbance, the UPS system immediately supports critical loads while standby generators are started and transferred onto the facility bus. Once generator operation is established, the facility transitions into an isolated backup operating mode until utility service is restored. Because backup resources are not continuously dispatched or coordinated with grid operations, the facility generally does not provide frequency support, voltage regulation, demand response, or other ancillary services \cite{nerc_datacenter_load_modeling}.

\subsubsection{Grid-Interactive Data Center Microgrid Architecture}

\begin{figure}[!h]
\centering
\includegraphics[width=\linewidth]{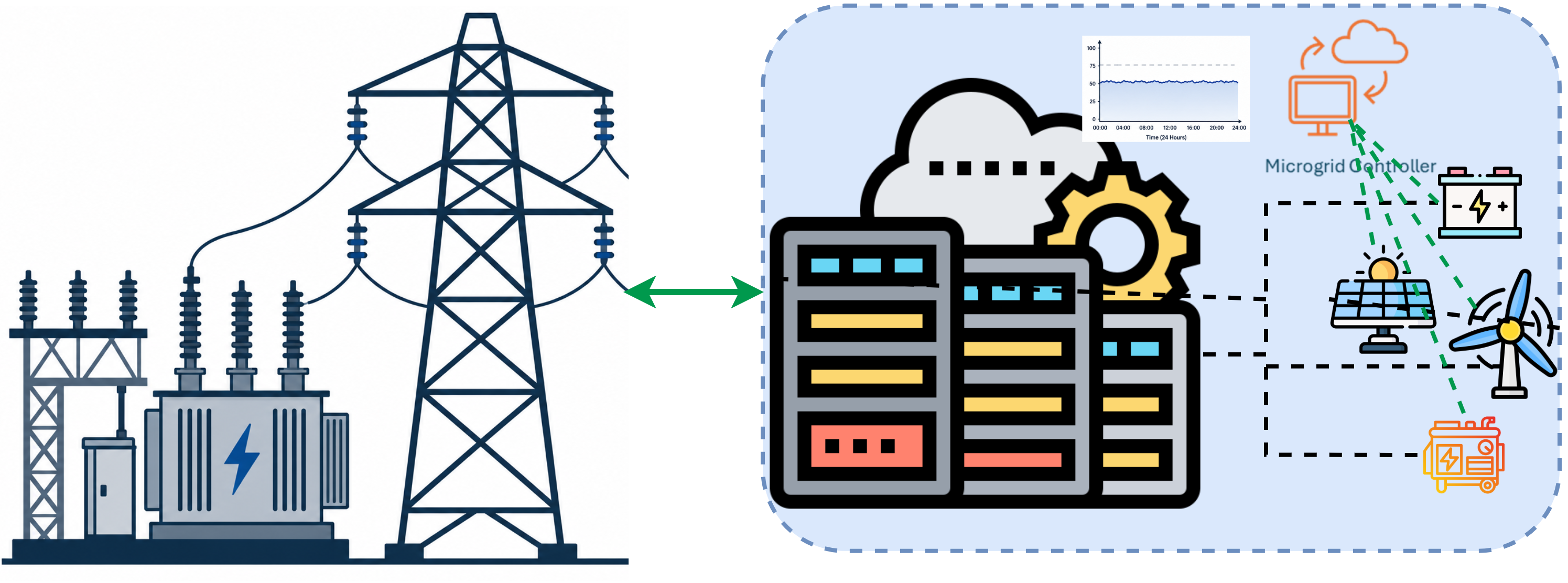}
\caption{Grid-interactive data center microgrid architecture.}
\label{fig:dc_microgrid_architecture}
\end{figure}

A grid-interactive data center microgrid integrates on-site generation, battery energy storage systems (BESS), IBRs, UPS systems, and intelligent supervisory controls with the utility grid through a point of common coupling (PCC), as illustrated in Fig.~\ref{fig:dc_microgrid_architecture}. Unlike conventional backup architectures, the microgrid controller continuously coordinates energy resources to optimize facility operation, improve resilience, and support grid reliability services.

The architecture consists of four primary layers: primary utility supply, microgrid resources, data center loads, and supervisory control. The key distinguishing characteristic of this architecture is that the facility can dynamically modify its net power consumption or injection in response to grid conditions. Through coordinated control of DGRs and BESS resources, the microgrid can perform peak shaving, demand response, frequency support, voltage regulation, and controlled islanding. Consequently, the data center behaves not only as a large load but also as an active grid participant capable of contributing to overall system reliability.

\subsection{Representation of Data Center Power Architectures}

A large portion of facility demand is supplied through uninterruptible power supplies (UPSs), rectifiers, and switch-mode power supplies that regulate their output and therefore behave approximately as constant-power loads near the nominal operating voltage. Cooling equipment, lighting, transformers, and other auxiliary systems introduce constant-current and constant-impedance characteristics. Consequently, the static portion of the data-center demand is represented using the ZIP polynomial model:

\begin{equation}
P(V)=P_0\left[
k_{pZ}\left(\frac{V}{V_0}\right)^2+
k_{pI}\left(\frac{V}{V_0}\right)+
k_{pP}
\right]
\label{eq:zip_p}
\end{equation}

\begin{equation}
Q(V)=Q_0\left[
k_{qZ}\left(\frac{V}{V_0}\right)^2+
k_{qI}\left(\frac{V}{V_0}\right)+
k_{qP}
\right]
\label{eq:zip_q}
\end{equation}

where $P_0$ and $Q_0$ are the active and reactive powers at the reference voltage $V_0$. The coefficients represent the constant-impedance ($Z$), constant-current ($I$), and constant-power ($P$) contributions and satisfy

\begin{equation}
k_{pZ}+k_{pI}+k_{pP}=1,
\qquad
k_{qZ}+k_{qI}+k_{qP}=1.
\label{eq:zip_constraints}
\end{equation}

In the absence of site-specific voltage-disturbance measurements, the coefficients listed in Table~\ref{tab:zip_base} are adopted as the base-case approximation.

\begin{table}[!h]
\caption{Base-Case ZIP Load Coefficients}
\label{tab:zip_base}
\centering
\begin{tabular}{lccc}
\hline
\textbf{Power Component} & \textbf{Constant} & \textbf{Constant} & \textbf{Constant} \\
& \textbf{Impedance} & \textbf{Current} & \textbf{Power} \\
\hline
Active Power ($P$)   & 0.10 & 0.10 & 0.80 \\
Reactive Power ($Q$) & 0.40 & 0.20 & 0.40 \\
\hline
\end{tabular}
\end{table}

The high constant-power fraction assigned to the active demand reflects the regulated behavior of UPS-fed computing equipment. The nonzero constant-current and constant-impedance fractions approximately account for transformers, lighting, auxiliary equipment, converter losses, and voltage-sensitive cooling demand. A larger impedance fraction is assigned to reactive power because passive magnetic equipment and fixed reactive components generally exhibit stronger voltage dependence.

These coefficients should not be interpreted as universal parameters for every data center. The facility composition depends on several factors, including the power usage effectiveness (PUE), UPS topology, cooling technology, power-factor correction strategy, server utilization, and the proportion of motors supplied through variable-frequency drives (VFDs). In particular, VFD-fed cooling systems appear to the grid primarily as power-electronic loads and should not automatically be represented as directly connected induction motors.

\section{Impact assessment of grid-interactive data center microgrids}
\label{sec:base-profile}
\begin{figure}[!h]
    \centering
    \includegraphics[width=0.8\columnwidth]{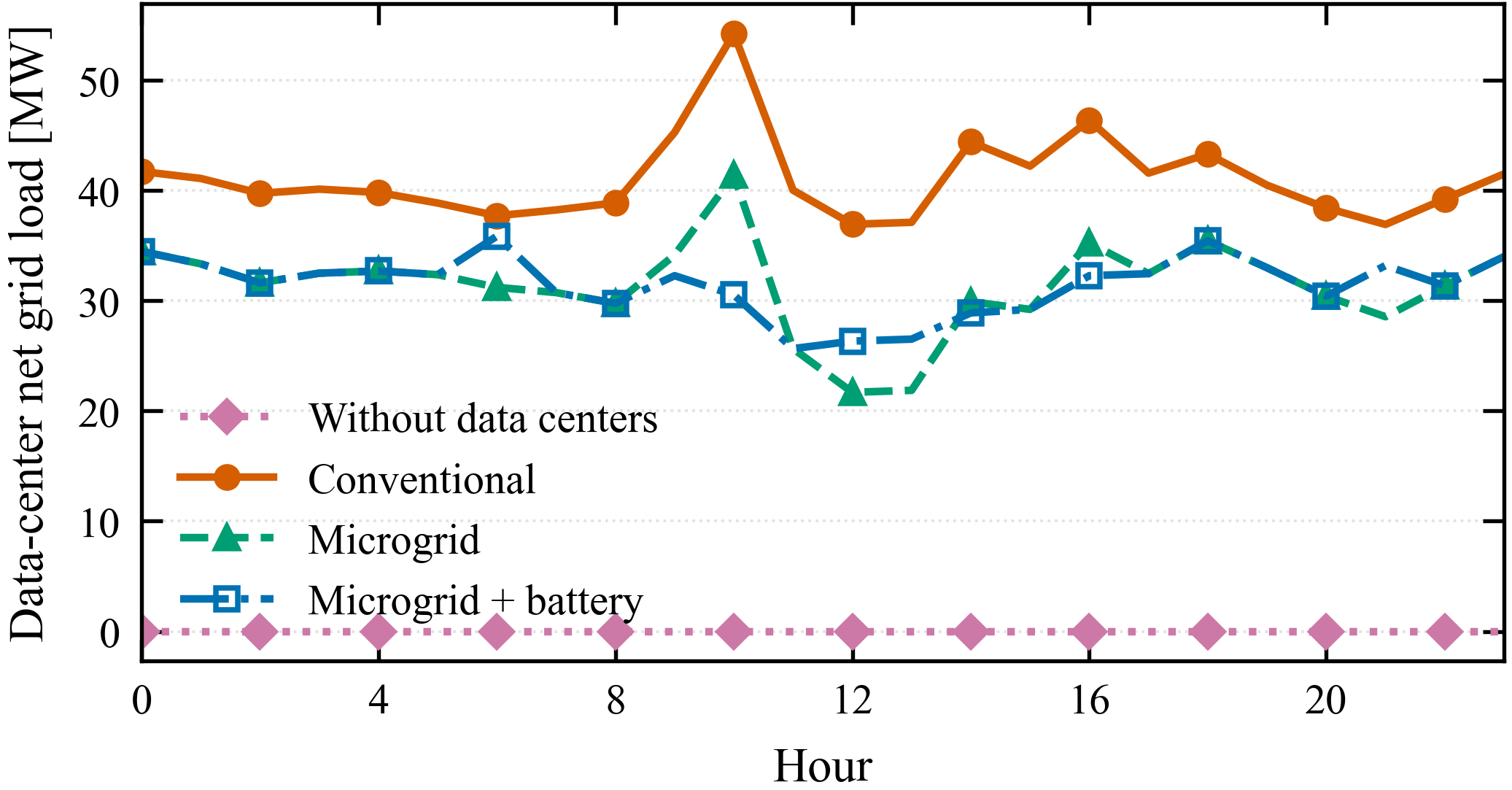}
    \caption{Data-center net grid import for the un-aligned peak.}
    \label{fig:dc-grid-import_a}
\end{figure}

\begin{figure}[!h]
    \centering
    \includegraphics[width=0.8\columnwidth]{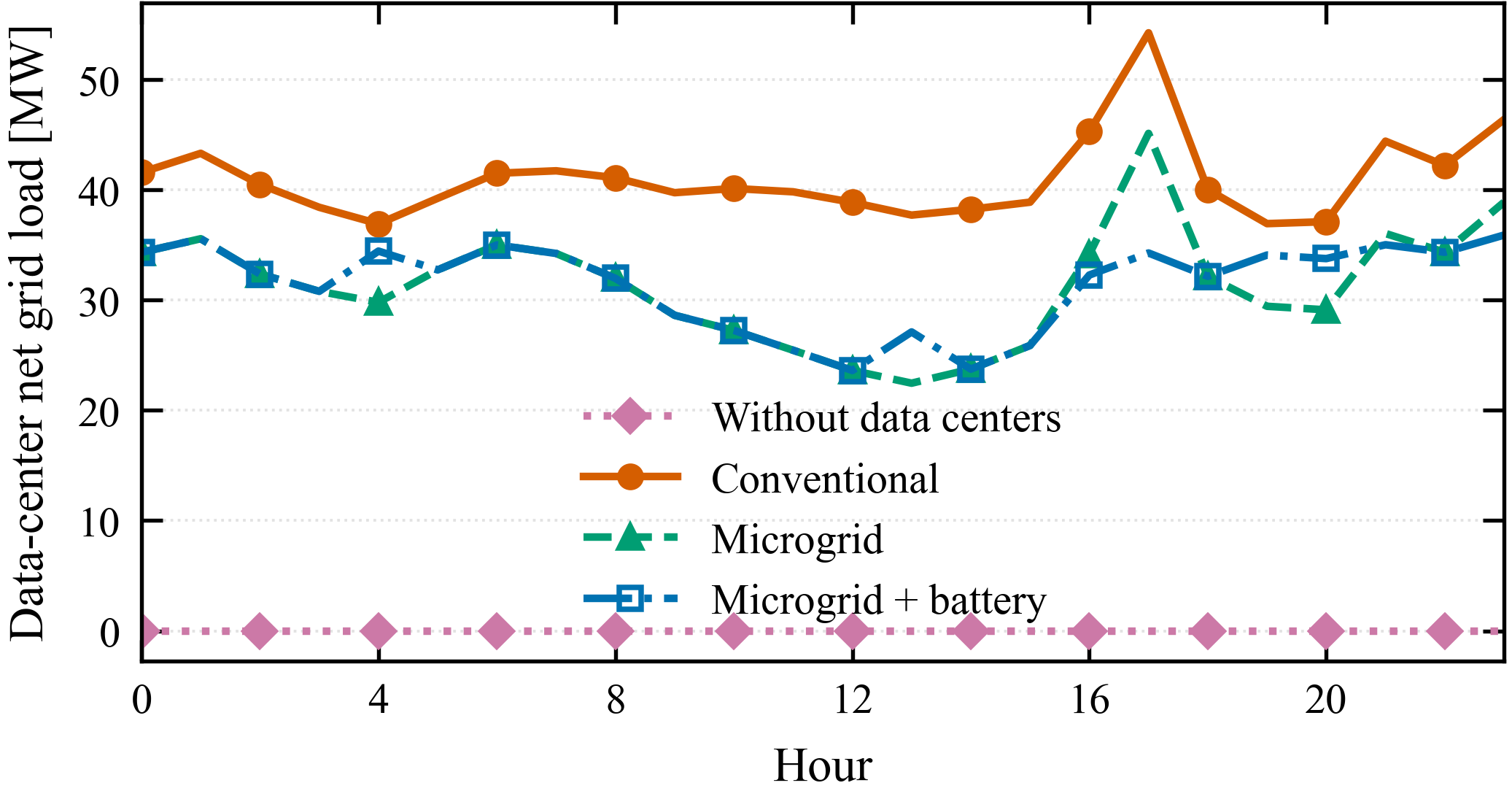}
    \caption{Data-center net grid import for the aligned-peak.}
    \label{fig:dc-grid-import_b}
\end{figure}

Impact of data center microgrids on the power system is assessed using the data centers included in the standard IEEE 14-bus system. This standard test system is analyzed with the daily load variations included. The original system contains \(223.7\) MW of active-power demand, which is scaled hourly using the system load profile. The scaling factor varies from \(0.62\) at hour 3 to \(1.15\) at hour 17. Consequently, the original system demand reaches its maximum value of \(257.26\) MW at hour 17. Using the scaling factors of each hour, effective loading on the test system is calculated. Two data centers are added to the system, with each facility having a rated capacity of \(25\) MW and operating at a lagging power factor of \(0.95\). Their active- and reactive-power demands are calculated as

\begin{align}
    P_{\mathrm{DC},i}(h)
    =
    P_{\mathrm{DC},i}^{\mathrm{rated}}\alpha_i(h), \\
    Q_{\mathrm{DC},i}(h)
    =
    P_{\mathrm{DC},i}(h)
    \tan!\left[\cos^{-1}(0.95)\right],
\end{align}

where $P_{\mathrm{DC},i}^{\mathrm{rated}}= 25\ MW$ and \(\alpha_i(h)\) is the facility-specific hourly load factor. The first data center is connected at bus 9 and the second at bus 14. Both buses belong to the \(138\) kV portion of the test system and are represented as PQ load buses. Before adding the data centers, bus 9 contains a native load of \(29.5\) MW and \(16.6\)MVar, whereas bus 14 contains \(20\) MW and \(7\) MVar. Bus 14 is located near the end of the network and is connected to buses 9 and 13. Its relatively weak electrical location makes bus 14 the limiting bus in the high-load and contingency studies. All these case studies are simulated using ANDES tool \cite{ANDES}. 
Three data center supply arrangements are compared. 
\begin{itemize}
    \item Conventional data center: the complete data center demand is supplied by the transmission system.
    \item Data center microgrid: each data center includes on-site generation, with each resource rated at \(25\%\) of the data center capacity. Thus, each facility has two inverter based resources (IBRs) of \(6.25\) MW each. The inverter-interfaced resources also compensate \(60\%\) of the data center reactive-power demand. The grid import is therefore
    \begin{equation}
        P_{\mathrm{grid},i}(h)=P_{\mathrm{DC},i}(h)-P_{\mathrm{on-site},i}(h) - P_{\mathrm{BESS},i}(h).
    \end{equation}
   The microgrid case without battery storage, \(P_{\mathrm{BESS},i}=0\).
    \item Battery-assisted data center microgrid: the battery is scheduled to shave \(20\%\) off the original data center peak. The required energy is replenished uniformly during four low-load hours using a charging efficiency of \(90\%\).
\end{itemize}
For all the three cases described above, two different operating conditions of the data center load pattern are analyzed. One is the base profile case where the data center loads peaks are not aligned with the system peak load and the other case is when peak loads of both data center and system load are aligned. To develop representative load profiles for electric distribution system studies, two co-located data centers were modeled using the SustainCluster simulation framework \cite{SustainCluster}. Each data center was configured with an IT capacity of 1~MW and located within the same geographic region (US-CAL-CISO), resulting in identical weather conditions, electricity price signals, carbon intensity data, and physical infrastructure characteristics. Independent workload queues and resource states were maintained for each facility to emulate operational diversity despite geographic co-location. Synthetic load profiles of the data centers for 1 day are presented in Fig. \ref{fig:dc-grid-import_a} and Fig. \ref{fig:dc-grid-import_b} respectively for un-aligned peak and aligned peak cases. It can be observed from Fig. \ref{fig:dc-grid-import_a} \& Fig. \ref{fig:dc-grid-import_b} that the minimum load for the data centers is $~0.5\ pu$ and the data centers experience high ramp rates of the compute loads which can change the load from 0.5 pu to 0.9 pu. 


\subsection{Un-aligned peak load case}
\begin{figure}[!h]
    \centering
    \includegraphics[width=0.8\columnwidth]{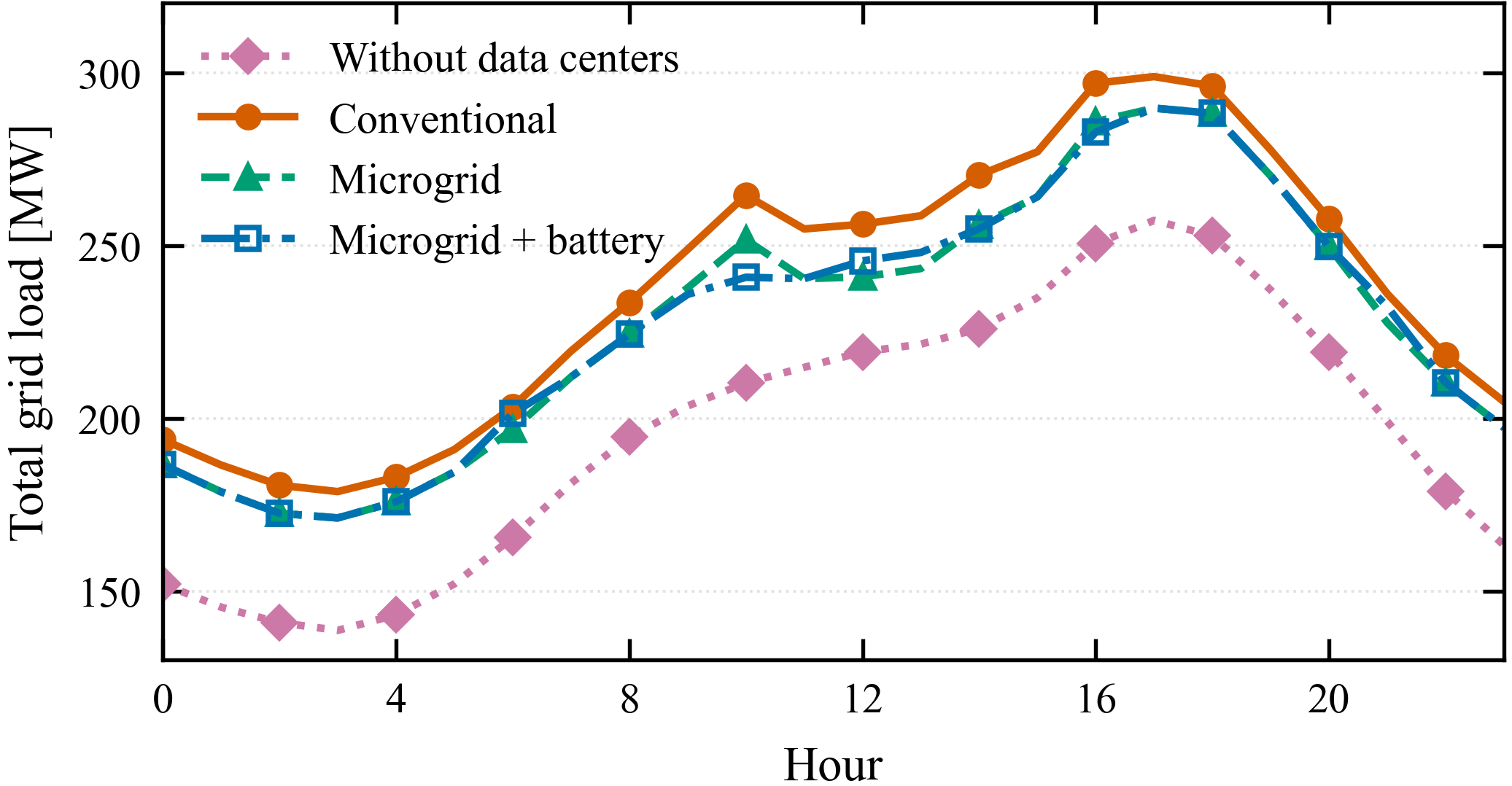}
    \caption{System load with data centers: un-aligned peak case.}
    \label{fig:total-grid-load_a}
\end{figure}

\begin{figure}[!h]
    \centering
    \includegraphics[width=0.8\columnwidth]{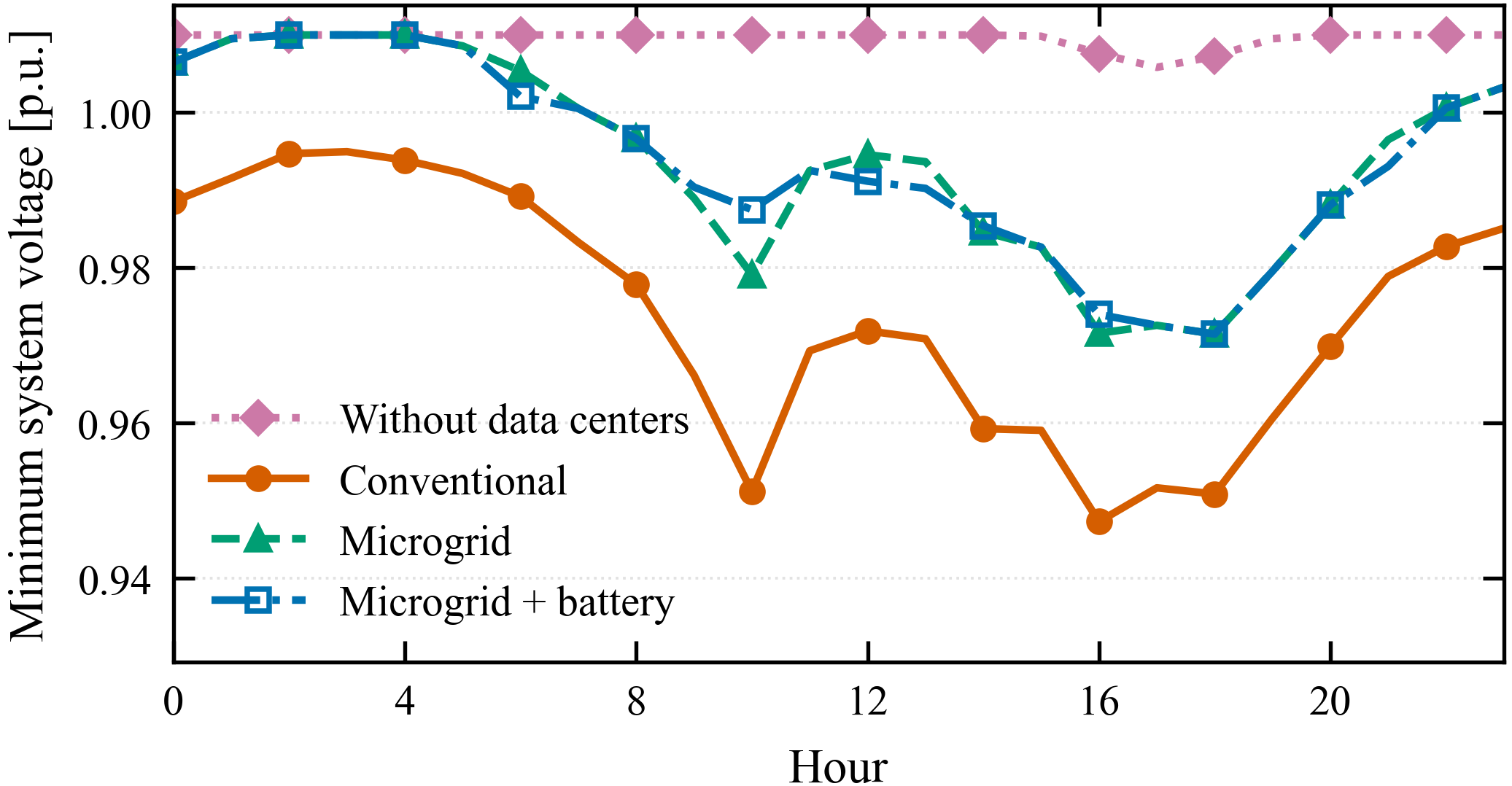}
    \caption{Minimum voltage over the study period: un-aligned peak.}
    \label{fig:min-voltage_a}
\end{figure}

In the base-profile case, the data-center load profiles are not synchronized with the system demand. Both data-center profiles attain their maximum values at hour 10, whereas the original system demand peaks at hour 17. At hour 10, the loads at buses 9 and 14 are \(24.66\) MW and \(29.59\) MW, respectively, giving a combined data-center demand of \(54.26\) MW. This temporal diversity reduces the coincidence between the data-center and system peaks.
Figure~\ref{fig:dc-grid-import_a} shows that the maximum data-center grid import is \(54.26\) MW for the conventional arrangement, \(41.38\)  MW for the microgrid arrangement, and \(35.87\) MW when battery storage is included. The battery increases the grid import slightly during its scheduled charging hours, but it limits the high data-center demand periods.
The total grid-load responses are presented in Fig.~\ref{fig:total-grid-load_a}. Although the data-center demand peaks at hour 10, the maximum combined system demand occurs at hour 17 because the original IEEE 14-bus load reaches its maximum at that hour. The corresponding total loads are \(298.84\) MW for the conventional arrangement and \(289.72\) MW for both microgrid arrangements. Battery discharge does not reduce the base-profile system peak because the battery schedule is determined by the earlier data-center peak rather than the hour-17 system peak.
Under normal operating conditions, the conventional arrangement produces a minimum voltage of \(0.9473\)~p.u. at hour 16. The microgrid and battery-assisted arrangements improve the minimum voltage to \(0.9715\)~p.u., as shown in Fig.~\ref{fig:min-voltage_a}. The improvement results from reduced active-power import and inverter-based reactive-power support. Post-contingency voltage security is quantified using two indicators: the minimum system bus voltage and the number of busses below the 0.95-p.u. voltage criterion.

\begin{figure}[!h]
    \centering
    \includegraphics[width=0.8\columnwidth]{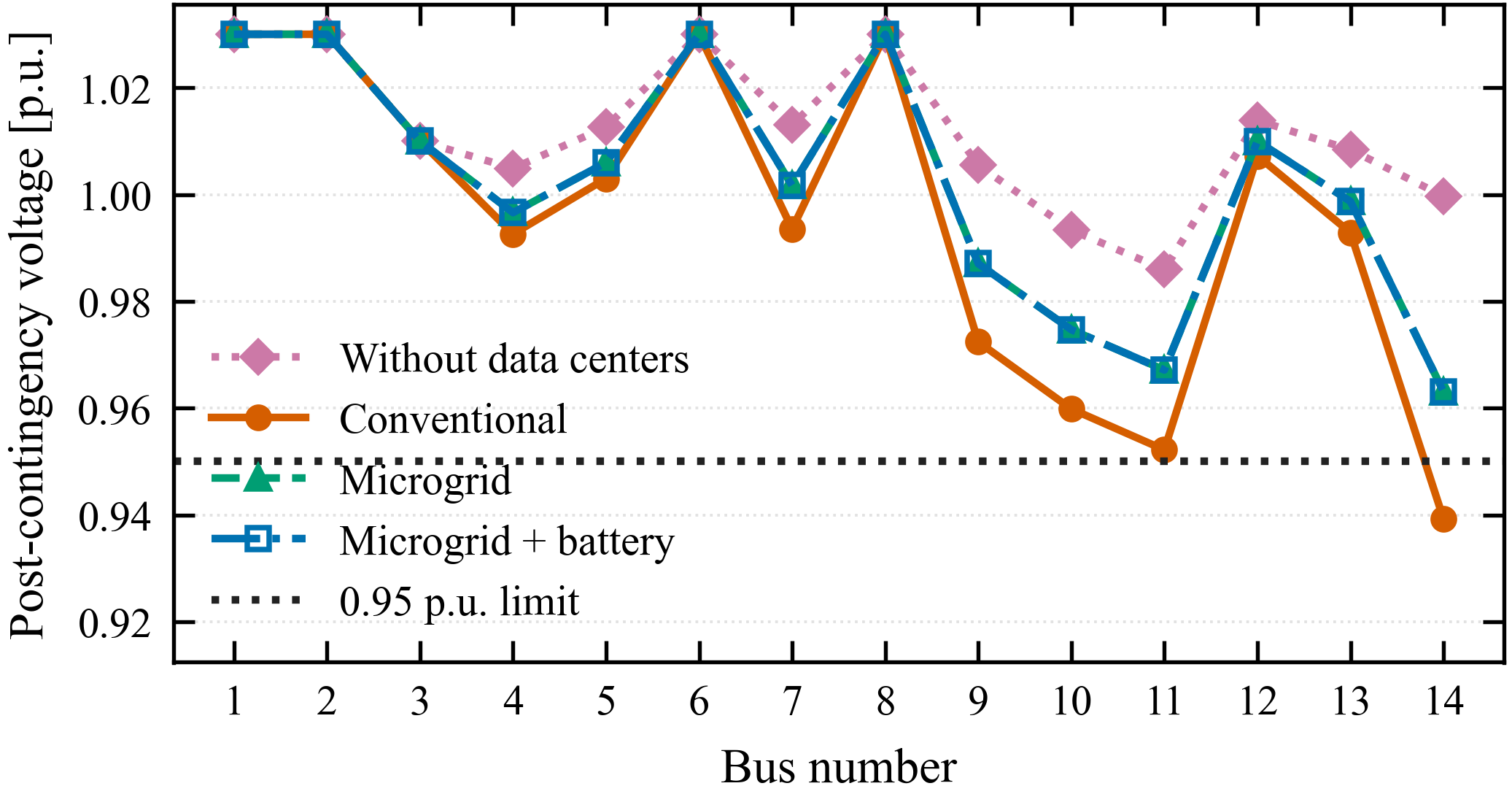}
    \caption{Post-contingency bus-voltage profiles following the outage
    of Line~8 for the un-aligned peak profile.}
    \label{fig:n1-voltage_a}
\end{figure}
The N--1 assessment removes Line~8, which connects buses 6 and 11, while retaining the hour-17 loading condition. As shown in Fig.~\ref{fig:n1-voltage_a}, the conventional case reaches a minimum post-contingency voltage of \(0.9393\)~p.u. at bus 14, with one bus below \(0.95\)~p.u. The microgrid and battery-assisted cases maintain a minimum voltage of \(0.9631\)~p.u., and no buses violate the \(0.95\)-p.u. criterion. The identical responses of the two microgrid cases again result from the battery being inactive at hour 17 in the base-profile schedule.
\subsection{Aligned peak load case}
\label{sec:aligned-profile}
\begin{figure}[!h]
    \centering
    \includegraphics[width=0.8\columnwidth]{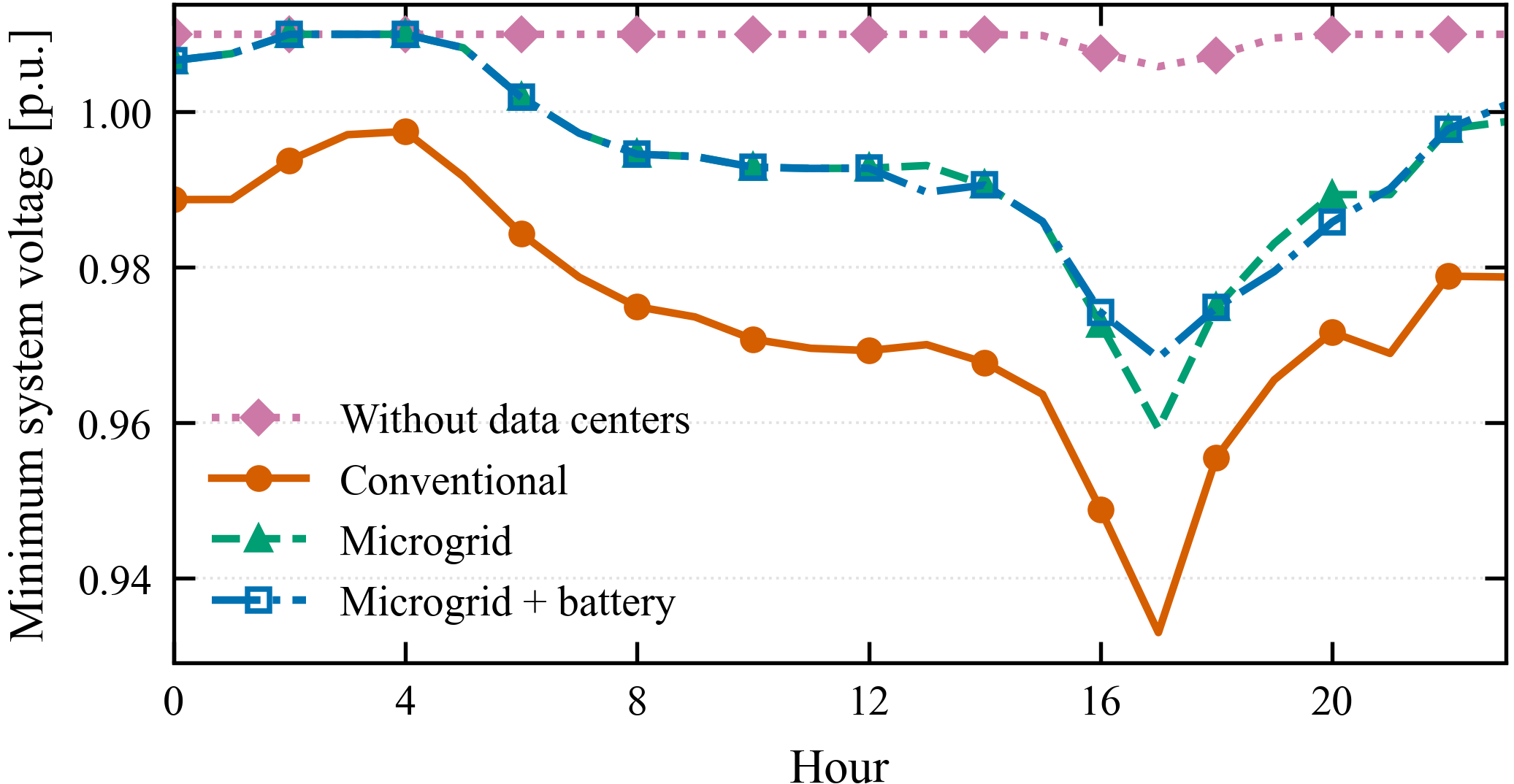}
    \caption{Minimum voltage over the study period: aligned peak.}
    \label{fig:min-voltage_b}
\end{figure}

The aligned-peak case uses the same IEEE 14-bus system, data-center capacities, connection buses, power factors, IBR capacities, reactive-power support, and battery parameters as the base-profile case. The only change is the timing of the two data-center load profiles. The original profiles are shifted so that their maximum demand occurs at hour 17, coincident with the \(1.15\)-p.u. maximum of the system load profile. The individual peak data-center demands remain \(24.66\) MW at bus 9 and \(29.59\) MW at bus 14, producing the same combined facility demand of \(54.26\) MW. Thus, the daily data-center energy is unchanged; only its temporal alignment with the system demand is modified.
The effect of peak alignment on data center grid import is shown in Fig.~\ref{fig:dc-grid-import_b}. At hour 17, the conventional data centers import the full \(54.26\) MW. Local solar and wind production reduce this value to \(45.13\) MW in the microgrid case. With battery storage, the hour-17 import is further reduced to \(34.28\) MW , corresponding to a battery discharge of approximately \(10.85\) MW . The maximum daily grid import of the battery-assisted arrangement is \(35.90\) MW and occurs at hour 23 rather than during the coincident system peak.
Peak alignment substantially increases the stress on the transmission system. The conventional total load reaches \(311.51\) MW at hour 17, compared with \(298.84\) MW in the base-profile case. IBR generation reduces the aligned-peak total load to \(302.39\) MW. Coordinated battery discharge further reduces it to \(291.53\) MW, which is approximately \(19.98\) MW below the conventional aligned-peak value.
The increased coincidence between the system and data center demands produces a more severe normal-condition voltage response. Figure~\ref{fig:min-voltage_b} shows that the conventional configuration reaches a minimum voltage of \(0.9330\)~p.u. at hour 17. This value is below the \(0.95\)-p.u. planning threshold. The microgrid improves the minimum voltage to \(0.9592\)~p.u., while the battery-assisted microgrid increases it further to \(0.9684\)~p.u. 

\begin{figure}[!h]
    \centering
    \includegraphics[width=0.8\columnwidth]{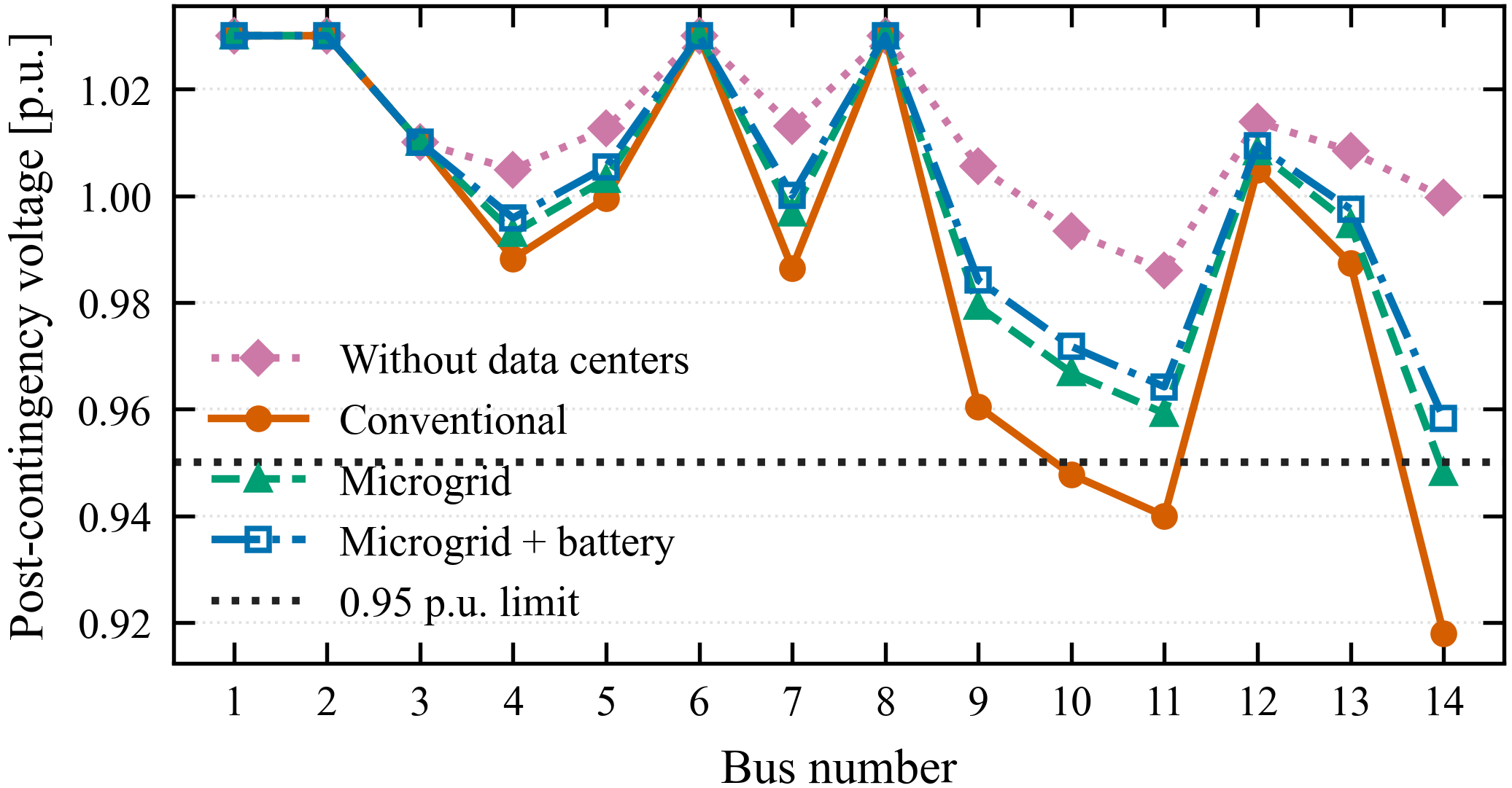}
    \caption{Post-contingency bus-voltage profiles following the outage
    of Line~8 for the aligned peak profile.}
    \label{fig:n1-voltage_b}
\end{figure}

The aligned-peak N--1 results are presented in Fig.~\ref{fig:n1-voltage_b}. Following the outage of Line~8, the conventional arrangement reaches a minimum voltage of \(0.9180\)~p.u. at bus 14, and three buses fall below \(0.95\)~p.u. The microgrid arrangement improves the minimum voltage to \(0.9482\)~p.u.; however, one bus remains below the voltage criterion. With battery storage, the minimum post-contingency voltage increases to \(0.9583\)~p.u., and all buses remain above \(0.95\)~p.u. Comparing the two cases isolates the effect of coincidence. Aligning the peaks raises the hour-17 system load by 12.67 MW in both the conventional and microgrid arrangements, while on-site generation provides the same 9.12 MW of relief in each. The battery, which discharges at the system peak only in the aligned case, removes a further 10.85 MW, most of the coincidence increase, and clears the remaining post-contingency violation. A battery scheduled to the facility's own peak therefore protects the system only when the peaks happen to coincide, which argues for scheduling data-center storage against system conditions.

\section{Conclusion}
This paper demonstrates that grid-interactive data center microgrids can transform large data centers from passive electrical loads into active grid assets. Comparative studies of conventional and grid-interactive architectures show that coordinated operation of on-site generation and battery energy storage can reduce peak grid demand and improve normal and post-contingency bus voltages. The results further highlight the importance of temporal alignment between data-center and system demand. Peak coincidence increases the system peak by approximately 12.7 MW, while on-site generation provides a similar reduction of approximately 9.1 MW regardless of peak alignment. However, under coincident peak conditions, on-site generation alone leaves a post-contingency voltage violation, whereas BESS discharge at the system peak eliminates the violation. These results demonstrate that data-center storage should be coordinated with system operating conditions rather than solely with the facility's own peak demand to maximize grid-level benefits. Overall, the comparative study quantifies the system-level benefits of grid-interactive data centers and highlights the importance of system-aware resource coordination for their reliable grid integration. Future work will focus on advanced control strategies, optimal resource coordination, economic assessment of grid services, and validation on larger transmission and transmission--distribution systems with multiple data-center interconnections.

\bibliographystyle{ieeetr}
\bibliography{ref.bib}

\end{document}